\documentstyle[prl,aps]{revtex}
\tightenlines

\newcommand{\be}{\begin{equation}}
\newcommand{\ee}{\end{equation}}
\newcommand{\bea}{\begin{eqnarray}}
\newcommand{\beas}{\begin{eqnarray*}}
\newcommand{\eea}{\end{eqnarray}}
\newcommand{\eeas}{\end{eqnarray*}} 
\newcommand{\ba}{\begin{array}}
\newcommand{\ea}{\end{array}}
\begin{document}

\draft
\twocolumn[\hsize\textwidth\columnwidth\hsize\csname
@twocolumnfalse\endcsname
\preprint{\vbox{
\hbox{UMD-PP-00-0xx}}}

\title{Inflation in Models with Large Extra Dimension Driven by a
Bulk Scalar Field}
\author{R. N. Mohapatra$^1$\footnote{e-mail:rmohapat@physics.umd.edu},
A. P\'erez-Lorenzana$^{1,2}$\footnote{e-mail:aplorenz@Glue.umd.edu} 
and C. A. de S. Pires$^{1}$\footnote{e-mail:cpires@physics.umd.edu}}
\address{
$^1$ Department of
Physics, University of Maryland, College Park, MD, 20742, USA\\
$^2$  Departamento de F\'\i sica,
Centro de Investigaci\'on y de Estudios Avanzados del I.P.N.\\
Apdo. Post. 14-740, 07000, M\'exico, D.F., M\'exico. }

\date{July, 2000}

\maketitle
\begin{abstract}
{We discuss inflation in models with large extra dimensions,
driven by a bulk scalar field.
The brane inflaton is then a single effective field, obtained from the
bulk scalar field by scaling. The self interaction terms of the effective
brane inflaton are then naturally suppressed. The picture is consistent
with a fundamental string scale in the TeV range without the problem of 
a superlight inflaton. 
If hybrid inflation
is considered, the right prediction for the density perturbations as
observed by COBE can be obtained without any fine tunning.
The bulk inflaton then decays preferentially into brane Higgses and 
reheating follows.} 

\end{abstract}
\vskip10pt]

%%%%%%%%%%%%%%%%%%%%%%%%%%%%%%%%%%%%%%%%%%%%%%%%%%%%%%%%%%%%%%%%%%%%%%
\section{Introduction}

Theories with extra dimensions where our four dimensional world is a
hypersurface (3-brane) embedded in a higher dimensional space (the bulk)
 have  been the focus of intense scrutiny during the last two years. 
It is generally assumed that in this picture the standard model particles
are in the brane whereas gravity and perhaps other standard model singlets
propagate in the bulk.
The main motivation for these models comes from string theories where the
Horava-Witten
solution~\cite{witen} of the non perturbative regime of the $E_8\times
E_8$ string theory provided one of the first models of this kind (although 
from a phenomenological point of view the idea was discussed early on by
several authors~\cite{rubakov,early}). Additional interest 
arose from the observation\cite{dvali} that the bulk size could be as
large as a millimeter leading to new observable 
deviations from Newton's inverse square law\cite{expt} at the millimeter
scale, where curiously enough Newton's law remains largely untested. 

A key formula that relates 
the string scale to the radius of the large extra dimension in these
models is:
 \be M^{2+n} R^n = M^2_{P\ell}, 
 \ee 
where $R$ is the common radius of the $n$ extra dimensions, $M$ is the
string scale and
$M_{P\ell}$ is the Planck scale. For $R\sim $ millimeter, $M$   can be as
low as few TeV thereby providing another resolution
of the long standing hierarchy problem. This has been another
motivation for these theories.

While this picture leads to many interesting consequences for collider and
other phenomenology\cite{exp}, it seems to require drastic rethinking of
the
prevailing view of cosmology \cite{riotto}. In particular, 
one runs into a great deal of
difficulty in implementing the standard pictures of
inflation\cite{inflation1}.
For instance, if the inflaton is required to be a brane field, its
mass becomes highly suppressed, making it difficult to understand
the reheating process. Also, a wall inflaton  makes it hard to understand
the density perturbations observed by COBE ~\cite{inflation1}.

As a way to solve these problems, Arkani-Hamed {\it et
al}~\cite{arkani-hamed}
proposed a scenario  where it was assumed that inflation occurs  before
the stabilization of the internal dimensions. With  the dilaton field 
playing the role of the inflaton field, they argued that early
inflation,
when the internal dimensions are small, can successfully  overcome the
above
complications. Another possible way out was proposed in Ref.~\cite{tye},
where it was suggested that the brane could be out of its stable point
at
early times, and inflation is induced on the  brane by its movement
through the extra space.  Still other ideas are found in~\cite{others}.

The common point of the first two scenarios is that they share the same
basic assumption of an unstable extra dimension. However, it is still
possible that, due to some dynamical mechanism, the extra dimension gets
stabilized long before the Universe exited from inflation, as in some
scenarios in 
Kaluza-Klein (KK) theories, where the stabilization potential is generated
by the Casimir force~\cite{kaluza-klein,tsujikawa}.
Other possible sources for this stabilizing potential could be present 
in brane-bulk theories; for instance, the formation of the brane at very
early times may give
rise to vacuum energy that plays a role in eventually 
stabilizing the extra dimension. 
It therefore appears to us that it would be of interest to
seek inflationary scenarios
where  stabilization occurs before inflation ends. Clearly in this case,
one cannot expect the dilaton field to play the role of the inflaton
field and we need
to find  a new way to generate inflation, that can solve the problems
 faced by the brane-inflaton.

With this background, we work in the framework of the
 Arkani-Hamed-Dimopoulos-Dvali (ADD) scenario where
stabilization of the internal dimensions occurred 
long before the end of inflation. The main new ingredient of our work is
that the inflaton is a bulk field rather than a field in the brane.
We give qualitative arguments to show that this provides a different 
way out to the problems introduced with a brane inflaton. 
We should stress that  while inflation  proceeds, in
principle, as in the former KK theories, the postinflationary era has a
 different behaviour 
mainly due to the fact that reheating must take place on the
boundaries
where all matter resides. This raises interesting questions regarding
the reheating process since, naively, one might expect
a bulk inflaton to  reheat the bulk instead
of the brane by releasing all its energy into the internal space in the
form of gravitons. However, as we will discuss, bulk heating is much
less efficient than brane heating, thereby circumventing this possibility.

The paper is organized as follows, in section 2, we review the problems
with brane inflation and comment on the possible solutions, leaving the
analysis of the present proposal (a bulk inflaton with stable bulk)  for
section 3. In  section  4, we discuss density perturbation. In
section 5, we address some details of the reheating era to explore the
puzzles introduced by the possible production of gravitons. 
We will close our discussion with some remarks.

\section{Brane inflation}

To see how letting the inflaton
arise from the brane fields
leads to problems~\cite{inflation1}, let us
consider a typical chaotic inflation scenario~\cite{chaotic}. If the
highest scale in the theory is $M$, during inflation, 
the inflaton potential  can not be larger than $M^4$, regardless of
the number of extra dimensions.
Since successful inflation (the slow roll condition) requires
that the inflaton mass be less than the Hubble parameter,  which is given
as 
 \be 
 H\sim \sqrt{{ V(\phi)\over 3 M_{P\ell}^2}}
 \label{eq2}
 \ee
we have the
inequality  
 \be 
 m\leq H\leq M^2/M_{P}.
 \label{suppres}
 \ee
For $M\sim 1$ TeV, one then gets the bound  $ m\leq 10^{-3}$ eV, which 
is a severe fine tuning constraint on the parameters of the theory.
It further implies that  
inflation occurs on a time scale $H^{-1}$ much
grater than $M^{-1}$. As emphasized by Kaloper and Linde~\cite{inflation1},
this is conceptually very problematic since it requires that 
the Universe should be large and homogeneous
enough from the very beginning so as to survive the large period of
time from $t=M^{-1}$ to $t=H^{-1}$.

Moreover, for chaotic inflation~\cite{chaotic} with 
$V(\phi) = {1\over 2} m^2 \phi^2$, we get for the density
perturbations  
 \be 
 {\delta\rho\over \rho}\sim 50 {m\over M_{P\ell}}\leq 10^{-31}.
 \ee
For the case where $\lambda\phi^4$ term dominates the density  one gets
 the
same old fine tunning condition ${\delta\rho\over \rho}\sim
\lambda^{1/2}$.
Assuming Hybrid inflation~\cite{hybrid}, with the potential
$V(\phi,\sigma)=
 {1\over 4\lambda} \left(M^2 - {\lambda} \sigma^2\right)^2 + 
 {1\over 2}m^2 \phi^2 + {g^2} \phi^2 \sigma^2 $,
does not improve those
results~\cite{inflation1}, since it needs either a value of $m$ six
orders of
magnitude smaller or a strong fine tunning on the parameters, to match
the COBE result ${\delta\rho\over \rho}\sim 10^{-5}$.

There are two possible ways to overcome this theorem. First, as emphasized
in Ref.~\cite{arkani-hamed}, we can imagine that during inflation era the
extra dimensions were as small as $M^{-1}$. Thus, instead of  Eq.
(\ref{eq2}) we will get 
 \be 
  H^2= {V(\phi)\over 3 M^2}; 
 \ee
which naturally removes the suppression (\ref{suppres}). However, one can
not allow the extra dimension to grow considerably during inflation since
large changes on the internal size will significantly affect
the scale invariance of the density
perturbations. Therefore the radius of the extra dimension must remain
essentially static while
the Universe expands. After inflation ends, the extra dimension should
grow to its final size, which may, however, produce a contraction period
on the
brane~\cite{arkani-hamed}. In this scenario, the radion should slow roll
during inflation and could be identified as the inflaton.
Nevertheless, it also poses some complications for the 
 understanding of  reheating
since the radion is long lived, and its mass could be very small (its mass
is lower bounded by $10^{-3}$ eV). 

Here, we consider the second possibility where the dynamics
of the  radion stabilizes the radius of the extra dimension before
inflation ends
($\tau_{stab}\ll\tau_{inf}$). There are examples of some KK inspired 
theories~\cite{kaluza-klein,tsujikawa} where this happens. In such a
theory, a different way around the
above problems is needed. Notice that, in this case, 
the scale invariance of density
perturbations requires that stabilization occurs  long before the last 80
e-foldings or so.
Clearly, in this case the radion can not play the
roll of the inflaton since it will not slow roll. 
As we will discuss in the next section,
if a bulk scalar field plays the role of the inflaton, a
simple solution to the above problems may be given.
One must however investigate the question of reheating carefully in this
model. Another important 
question in this model is the origin of stabilization of the extra
dimension. We do not  
address this difficult question here but simply assume
the condition that $\tau_{stab}\ll\tau_{inf}$.

\section{Inflation with a bulk scalar field}

Let us now discuss the picture of inflation, when the inflaton is
 a higher dimensional scalar field.  To keep things simple, we will
assume only a single extra dimension. However, we stress that our results
hold for any number of extra dimensions as long as the inflaton propagates
in all of them.

Let us start by assuming  that the extra dimensions are  already
stabilized by some (yet unknown) dynamical mechanism~\cite{radion}. As the
inflaton is now a bulk field, we will further assume that it is
homogeneous along the extra dimensions, just as in the former KK theories.
This is another way to state the 
perfect fluid assumption
for the $\Phi$ field in five dimensions (i.e. $T_{05}=0$ where $T_{05}$ is
one of the components of the energy momentum tensor).
Obviously this will make our theory of inflation similar to those models.
However, what will make our theory different from the usual KK theories 
is the fact that matter is  attached to the branes and that will affect
the inflaton decay in an essential manner.
Notice that the condition $T_{05}=0$ makes the inflaton  a  zero mode,
which is also a necessary condition if we want to reproduce
the ADD scenario at late times (a flat and factorizable geometry). 
Therefore, in the effective four dimensional theory, the inflaton field
$\tilde\phi$ and the bulk field $\Phi$ are  related by
$\tilde\phi=\sqrt{R}\Phi_0$. 
Notice that this assumptions are consistent as long as the brane densities
remain smaller than $M^4$. If brane densities were large, one
would have to consider the branes as sources for the metric
in the Einstein equations and one will depart from the ADD picture towards
a Randall-Sundrum type nonfactorizable geometry. We will 
not consider such scenarios here. Some ideas on this regard can be found
in Ref.~\cite{branecosmology}. 

Once the extra dimension is stable, one gets, in the effective four
dimensional theory, the usual form of the Hubble parameter as
\be 
  H^2 = {V_{eff} \over 3 M^2_{Pl}};
  \label{H2}
 \ee
where the effective four dimensional potential is defined by
\be 
 V_{eff} = R V_{5D} = 
 \left({M^2_{P\ell}\over M^3}\right) V_{5D} .
 \ee
As we will see, this condition translates into the scaling of the inflaton
as mentioned above. Now we focus on the implications of this formula.

First, since the inflaton is now a bulk field, the upper bound
on the five dimensional  potential is $M^5$ (instead of $M^4$ for the
case of the brane inflaton) and the effective 
potential has the upper bound $R M^5 = M^2M_{P\ell}^2$. Therefore, one
gets $m\leq H\leq M$, which does not require a superlight inflaton. This
also keeps the  explanation of the flatness and horizon problems as
usual, since
now, the time for inflation could be as short as in the standard theory.
Thus, our bulk field naturally overcome the problem noted by Lyth and 
Kaloper and Linde.

To proceed further, let us
assume the following five dimensional potential for the bulk field:
 \be 
  V_{5D}(\Phi) = \frac{1}{2} m^2 \Phi^2 + \frac{\lambda}{4 M} \Phi^4.
 \label{vphi}
  \ee
The effective potential that drives inflation on the brane can be derived
from the above equation to be
 \be 
 V_{eff}(\tilde\phi) = \frac{1}{2} m^2 (\tilde\phi)^2 + 
 \frac{\tilde\lambda}{4} (\tilde\phi)^4.
 \label{veff}
  \ee
where $\tilde\lambda := {M^2\over M_{P\ell}^2}\lambda$ is a naturally 
suppressed coupling. Now, as in the old fashioned chaotic
scenario, inflation will start in those small patches of size $H^{-1}$
where the effective inflaton reaches an homogeneous value 
$\tilde\phi_c\geq M_{P\ell}$. However, because of scaling,
this requires that for the bulk field, we must have $\Phi_c(0)\geq
M^{3/2}$, which is a natural value in our picture.

We wish to note parenthetically,
 that if the $\Phi^4$ term dominates the energy density (i.e. $m\ll
M$), then $\tilde\phi$ could develope a vacuum expectation value in which 
case an interesting connection between the scales $M,~m$ and $v_{ew}$ can 
emerge as follows. We can have the inflaton couple to  matter fields
in the brane (which is needed to reheat the brane Universe), via the
following term:
 \be 
   h M^{1\over 2}\Phi \chi^2 \delta(y),
 \ee
where $h$ is a dimensionless coupling constant and $\chi$ is the brane
higgs field.
When $\Phi$ develops a vacuum expectation value, the last term will
contribute to the
mass term, $\mu^2_0$, in the Higgs potential, which should be of order of
the weak scale. From the potential given in
(\ref{vphi}), we get then the constraint
 \be 
  \mu_0^2 = \frac{h}{\lambda^{1\over 2}}m M \approx v^2_{ew}.
  \label{mu0}
  \ee
  Now assuming that $h,\lambda\sim 1$ and $m\sim 10$ GeV $\ll M$,
 we get $M\sim 10^2\ TeV$,
which is consistent with the strongest experimental limit~\cite{exp}.

%%%%%%%%%%%%%%%%%%%%%%%%%%%%%%%%%%%
\section{Density perturbation}

The calculation of the density perturbation proceeds as in  the usual
four dimensional theories. Let us write it down 
 in terms of the five dimensional potential:
 \be 
  {\delta\rho\over \rho} \sim 
  {\left(V_{eff}(\tilde\phi_c)\right)^{3/2} \over 
  M_{P\ell}^3 \left({\partial V_{eff}\over\partial\tilde\phi_c}\right)} 
  =  \left(1\over M_{P\ell} M^3\right)
   {\left(V_{5D}(\Phi_c)\right)^{3/2} \over  
   \left({\partial V_{5D}\over \partial\Phi_c}\right)}.
  \label{drho} 
 \ee
Because the quartic term is  suppressed for small values of $M$, we first
assume that the mass term drives the inflation. 
Nevertheless, as expected we get 
${\delta\rho\over \rho} \sim m/M_{P\ell}$, which is again very small 
for $m \leq M$. This result is similar to what one obtains in the brane
inflaton models\cite{inflation1}. On the other hand, if the quartic term
dominates the density, then 
 \[ 
 {\delta\rho\over \rho} \sim \tilde\lambda^{1/2} =
 \left({M\over M_{P\ell}}\right) \lambda^{1/2}.
 \]
and models with only large values of $M$ would be satisfactory.
Since our interest here is in models with large extra dimensions, we
consider $M$ in the multi-TeV range and therefore we must seek ways to
solve this problem. In any case it is gratifying that a single bulk scalar
field seems to solve two of the major problems faced by the brane inflaton
models.

In order to improve the situation with respect to $\frac{\delta
\rho}{\rho}$ in this model, we extend it to
include an extra scalar field $\sigma$ and
 considering bulk potential to have the same form as is used in
implementing hybrid inflation~\cite{chaotic} picture:
 \be
 V(\phi,\sigma) = {M\over 4\lambda}
 \left(M^2 - {\lambda\over M} \sigma^2\right)^2 + 
 {m^2_0\over 2} \phi^2 + {g^2\over M} \phi^2 \sigma^2 .
 \ee
It is easy to check that inflation will require
$\phi^2_c\geq M^3/2g^2$. Therefore, our effective inflaton 
should be $\tilde\phi_c\geq
M_{P\ell}/\sqrt{2}g$, just as expected. One then uses (\ref{drho}) to get
 \be 
  {\delta\rho\over \rho} \sim 
  \left({g\over 2 \lambda^{3/2}}\right)\ {M^3\over m_0^2 M_{P\ell}}.
 \label{drhyb}
 \ee
If for instance, we set in the last equation the values
 $M\sim 10^2\ TeV$ and $m_0\sim m\sim 10\ GeV$ we find
 \be
 {\delta\rho\over \rho} \sim 
  \left({g\over 2 \lambda^{3/2}}\right)\times 10^{-5},
 \ee
which is  right the COBE result. 

Let us now compare our model with the case where one has hybrid
inflation in the wall. In order to explain
the density pertubation in the wall hybrid inflation models , an 
unpleasant fine tuning of either the the mass
of the inflaton field, (to the level of $m_0=10^{-10} eV$), or  of the
coupling constant, (to the level of
$\lambda=10^{-8}$\cite{inflation1}) is essential. On the other hand as we
just showed, if the inflaton is a bulk field, no such fine
tuning is required. We find this to be perhaps an interesting
advantage of models with large extra dimensions over the conventional four
dimensional inflation models.

Let us also 
remark that in the case of more than one extra dimension, the only
change on our above results come form the substitution of $R$ for the
volume of the extra space $V_n$ in all the analysis. This does not affect
the results of our analysis, since the effective theory is still given in
terms of
the same effective coupling constants, although the effective inflaton will
be changed into $\sqrt{V_n} \Phi$. This rescaling does not affect our
main expression in (\ref{drho}) nor  (\ref{drhyb})

%%%%%%%%%%%%%%%%%%%%%%%%%%%%%%%%%%%%%%%%%%%%%%%%%%%%%%%%%%%%%%%%%%%%%%%%%%%%%%%%%%
\section{Reheating}

The epoch of the Universe soon after inflation is called reheating. During
this era, the inflaton is supposed to decay into matter populating
the Universe and reheating it to a temperature $T_{R}$, called
the reheating temperature.
Since many important phenomena of cosmology e.g. baryogenesis, depend on
the Universe being very hot, the value of $T_R$ is important.
One also has to watch out for any unwanted particles that
may be produced during the reheating, since they may create problems
for the subsequent evolution of the Universe (as for instance is
familiar from the study of gravitino production in supergravity theories). 
Reheating is followed by thermalization of particles
produced, so that conventional Friedman expansion can begin subsequently.
Again, thermalization is also dependent on $T_R$. Clearly, therefore
reheating is a very important aspect of any model of inflation. In this
section, we discuss how it works in our model.

In our discussion, we will use the elementary theory of reheating, which 
is based on perturbation
theory \cite{reheating1} and where the reheating temperature
 $T_R$ can be expressed in terms of the total decay rate of the
inflaton, $\Gamma$ and the Hubble parameter as follows: $T_R \sim
0.1\sqrt{\Gamma M_{P\ell}}$ \cite{reheating2}.
It has been pointed out that this approach  
faces some limitations in terms of efficiency\cite{reheating2} and
possible improvements have been suggested using a first
stage of heating (called preheating\cite{preheating}) based on parametric
resonance. We will not be concerned here with these extra subtleties and
use
perturbative reheating to get a crude idea of the postinflationary
phase of the Universe and discuss how the Friedman Universe emerges
following the end of the reheating period.

An obvious problem that could possibly arise is that 
the strong coupling of the KK modes of graviton, could induce a faster
decay
channel for the inflaton than matter. If this happens, the bulk would
reheated while the brane would remain empty of matter
giving rise to a non-standard, undesirable Universe.
As we show below, luckily this is not the case for our model.

When hybrid inflation ends, the field $\sigma$ quickly goes to one of its
minima $\sigma_{\pm}=\pm M^{3/2}/\sqrt{\lambda}$. As a  result the
mass term of the inflaton field receives a contribution
$\sim g^2M^2/\lambda$
which  dominates over $m$. This leads us to conclude that the
inflaton field, which will oscillate around its minimum and generate
reheating, has a mass about an order of magnitude below $M$ 
(using the set of parameter chosen to explain
density perturbations). Therefore, the decay $\phi\rightarrow \chi\chi$ is
allowed, where $\chi$ is the Higgs field, for typical Higgs field masses
in the 100-200 GeV range. Let us stress that
this process will take place only on the boundaries of the extra
dimension (i.e. in the brane),  making it physically
different from the former KK theories where the production of matter
through inflaton decays occurs everywhere.

Following the steps of perturbative reheating theory, we can estimate  the
reheating  temperature by calculating the decay rate 
of the inflaton field into
two Higgses. That is 
 \be 
 \Gamma_{\phi\rightarrow \chi \chi}\sim 
 {M^4\over 32 \pi M_{P\ell}^2 m_\phi}   .
 \label{g1}
 \ee
With an inflaton mass $m_\phi$
around $0.1 M$  we  estimate the reheating temperature to be $T_R > 100$
MeV. As would be desirable, this temperature is above that required for
successful big bang nucleosynthesis.

The next point that needs to be investigated is the generation of
gravitons by the excited modes of  $\phi$  and $\sigma$ fields.
Because, if excessive graviton production drained away the energy
stored in the inflaton field, it would lead to lower matter density
compared to graviton density and matter may not reach a state of
equilibrium. In the five dimensional language, this would lead to an
expansion of the bulk rather than the brane.
Notice that such processes are not dangerous in KK theories where the
compactification scale is very small and the excited modes decay
preferentially into matter (on the bulk).
 Now, that the radius of the extra space is large, a large number
of KK gravitons, $h$, could be produced by a KK inflaton mode
decaying into another excited mode plus a graviton: 
$\phi_n \rightarrow \phi_l h_{n-l}$, where $n$ and $l$ are the
KK numbers, which are conserved. To estimate the amount of gravitons
produced, we have to first estimate the production rate for $\phi_n$, the
excited modes of the inflaton, since only $\phi_n$
 decay can produce gravitons via the decay process just mentioned.
The $\phi_n$ modes are produced through via collision processes, 
$\phi_0 \phi_0  \rightarrow \phi_n \phi_n$ and the rate for this process 
is given by 
 \be 
\sigma_{\phi_0 \phi_0\rightarrow
\phi_n \phi_n} \sim \lambda^2 \frac{M^2}{M_{Pl}^4}; 
 \ee
for $\sqrt{s}\sim M$. 
This  has to be compared with $\Gamma_{\phi\rightarrow \chi \chi}$ above
(\ref{g1}). Due to the very different Planck mass dependence, it is easy
to see that the $\phi_n$ production is highly suppressed compared to the
$\phi_0$ decay to Higgs bosons. Once an excited mode has been
produced, it will preferably  decay into gravitons.
Although the rate for this  procces is very small:
 \be 
 \Gamma_{\phi_n\rightarrow \phi_l h_{n-l}}\sim 
 {m_n m_l^2\over 12 \pi M_{P\ell}^2 }   ; 
 \ee
where $m_n^2= m_\phi^2 + n^2/R^2$ is the  mass of the excited mode,
the presence of a large number of accessible modes in the final
state will enhance this value up to 
  \be 
 \Gamma_{\phi_n, Total}= 
 \sum_l \Gamma_{\phi_n\rightarrow \phi_l h_{n-l}}\sim 
 {m_n^3\over 12 \pi M^2 }   ; 
 \label{g2}
 \ee
making the excited mode very short lived. We should stress, however, that
the final products of the shower induced by a KK mode decaying on 
lighter modes will always include $\phi_0$'s, which as we stated already
only decay on the brane to Higgs fields and hence to matter. As a result,
the KK excitations of the $\phi_0$ will not be around to overclose the
Universe and the associated graviton production is also unlikely to be
significant.

Thus, the final scenario that emerges is as follows:  after exiting
inflation
the inflaton will start moving relativistically, eventually producing
both Higgs fields as well as its excited modes due to the
possible (suppressed) four points self interactions. As the rate estimates
above show, most of the reheat energy will pass to the brane in the form
of matter and a very small part will pass to the bulk in the
form of gravitons produced through the fast decaying KK inflaton modes. As
the Universe is still rapidly 
moving at that stage, we could imagine that the
density of those gravitons will be substantially diluted. The 
 Higgs bosons produced by the inflaton decay will quickly decay to quarks
and leptons, which will attain equilibrium, via their strong and weak
interactions. Friedman
expansion will resume, albeit starting with a lower temperature ($T\sim
0.1 $ GeV) compared to the conventional grand unified theories.

We also point out that this scenario, naively, will not be affected
by the presence of
the $\sigma$ field of the hybrid inflation model.

%%%%%%%%%%%%%%%%%%%%%%%%%%%%
\section{Remarks and discussion}

We now conclude with a brief summary of our main results.  Choosing the
bulk scalar field as the source of the brane inflaton field leads to
several advantages: first the fundamental scale
$M$ can be in the multi-TeV range, which turns out to be the natural
bound for inflaton mass $m_{\phi}$ and  Hubble constant $H$, in
contrast with the brane
inflaton models where $m_{\phi}$ and $H$ are oversuppressed. Assuming
hybrid inflation
and $M$ just above the current experimental limits,  the COBE observation
of $\delta\rho\over\rho$ is also successfully  explained without any fine
tuning. Finally, we also note that even though the bulk inflaton has KK
modes, the reheating process leads to Universe not dominated by their mass
but rather by the standard model particles in equilibrium.
Finally, we mention that all the results of this paper remain unchanged 
when more extra dimensions are involved, provided that the inflaton
propagates in all the bulk, and that the Friedmann equation (Eq.
(2)) holds.

%%%%%%%%%%%%%%%%%%%%%%%%%%%%%%

{\it Acknowledgements.}   The work of RNM is supported by a grant from the
National Science Foundation under grant number PHY-9802551. The work of
APL is supported in part by CONACyT (M\'exico). The work of CP is
supported by Funda\c c\~ao de Amparo \`a Pesquisa do Estado de S\~ao Paulo
(FAPESP). We like to thank S. Nussinov for several stimulating discussions
on the reheating process.

%%%%%%%%%%%%%%%%%%%%%%%%%%%%%%


\begin{thebibliography}{99}

\bibitem{witen}
E. Witten, Nucl. Phys. {\bf B471}, 135 (1996);
 P. Horava and E. Witten, Nucl. Phys. {\bf B460}, 506 (1996);
 {\it idem} {\bf B475}, 94 (1996).


\bibitem{rubakov}
V. A. Rubakov and M. E. Shaposhnikov,  Phys. Lett. {\bf B152}, 136 (1983);
K. Akama,  in Lecture Notes in Physics, 176, Gauge
 Theory and Gravitation, Proceedings of the International Symposium on
 Gauge Theory and Gravitation, Nara, Japan, August, 20 (1982), edited by
 K. Kikkawa, N. Nakanishi and H. Nariai, (Springer-Verlag), 267 (1983);
M. Visser, Phys. Lett {\bf B159}, 22 (1985);
E. J. Squires, Phys. Lett {\bf B167}, 286 (1986);
G. W. Gibbons and D. L. Wiltshire, Nucl. Phys. {\bf B287}, 717 (1987).

\bibitem{early}
For some early ideas of a low string scale see also:
I. Antoniadis, Phys. Lett. {\bf B246}, 377 (1990); 
I. Antoniadis, K. Benakli and M. Quir\'os, 
Phys. Lett. {\bf B331}, 313 (1994);
K. Benakli, Phys. Rev. D{\bf 60}, 104002 (1999); \pl {\bf B447}, 51 (1999).

\bibitem{dvali}
N. Arkani-Hamed, S. Dimopoulos and  G. Dvali, \pl {\bf B429}, 263 (1998); 
 \prd {\bf 59}, 086004 (1999);
I. Antoniadis, S. Dimopoulos and G. Dvali, 
Nucl. Phys. {\bf B516}, 70 (1998); 
N. Arkani-Hamed, S. Dimopoulos and J. March-Russell, hep-th/9809124.

\bibitem{expt} 
J. C. Long, H. W. Chan and J. Price, Nucl. Phys. {\bf B539}, 23 (1999).   

\bibitem{exp}
For  experimental bounds see for instance: 
L. J. Hall and  D. Smith. Phys. Rev. D {\bf 60}, 085008 (1999);
T. G. Rizzo, Phys. Rev. D {\bf 59}, 115010 (1999);
G. F. Giudice, R. Rattazzi and J. D. Wells, 
  Nucl. Phys. {\bf B544}, 3 (1999);
E. A. Mirabelli, M. Perelstein and Michael E. Peskin, 
Phys. Rev. Lett. {\bf 82}, 2236 (1999);
J. L. Hewett,  Phys. Rev. Lett. {\bf 82}, 4765 (1999);
V. Barger, T. Han, C. Kao and R. J. Zhang, \pl {\bf B461}, 34 (1999).

\bibitem{riotto}
For a review, see D. H. Lyth and A. Riotto, 
Phys. Rept. {\bf 314}, 1 (1999);


\bibitem{inflation1}
D. H. Lyth, Phys. Lett. {\bf B448}, 191 (1999);
N. Kaloper and A. Linde, Phys. Rev. D {\bf 59}, 101303 (1999).

\bibitem{arkani-hamed}
N. Arkani-Hamed, {\it et al.},
Nucl. Phys. {\bf B567}, 189 (2000).

\bibitem{tye}
G. Dvali, S. H. H. Tye, Phys. Lett. {\bf B450}, 72 (1999).

\bibitem{others}
For some other references about inflation in models with extra dimensions
see for instance:
D. H. Lyth, Phys. Lett. {\bf B466}, 85 (1999);
E. Halyo, Phys. Lett. {\bf B461}, 109 (1999);
A. Mazumdar, Phys. Lett. {\bf B469}, 5 (1999);
A. Mazumdar, Jing Wang; gr-qc/0004030; A. Mazumdar, hep-ph/0008087; 
T. Nihei, Phys. Lett. {\bf B465}, 81 (1999);
R. Maartens, D. Wands, B. A. Bassett and I. Heard, 
Phys. Rev. D{\bf 62}, 041301 (2000);
E. J. Copeland, A. R. Liddle, J. E. Lidsey; astro-ph/0006421;
L. E. Mendes, A. R. Liddle; astro-ph/0006020;
P. Kanti and K.A. Olive, Phys. Rev. D {\bf 60}, 043502 (1999).


\bibitem{kaluza-klein}
L. Amendola, E. W. Kolb, M. Litterio, F. Occhionero,
 Phys. Rev. D{\bf 42}, 1944 (1990).
 
\bibitem{tsujikawa} 
 S. Tsujikawa, hep-ph/0005105.
 
 \bibitem{chaotic}
A. Linde, \pl {\bf B129}, 177 (1983).

\bibitem{hybrid}
A. Linde, \pl {\bf B259}, 38 (1991); \prd {\bf 49}, 748 (1994).

\bibitem{radion}
For some ideas on the  dynamics of the stabilization 
of the extra dimension and inflation  see for instance:
C. Csaki, M. Graesser and  J. Terning, Phys. Lett. {\bf B456},  16 (1999);
J. M. Cline, Phys. Rev. D{\bf 61}, 023513 (2000);
E. E. Flanagan, S. H. H. Tye and  I. Wasserman,  hep-ph/9909373; 
A. Riotto, Phys. Rev. D{\bf 61}, 123506 (2000);
W. D. Goldberger, and M. B. Wise; Phys. Rev. Lett.{\bf 83}, 4922 (1999).

\bibitem{branecosmology}
See for instance
P. Binetruy, C. Deffayet and D. Langlois, 
Nucl. Phys. {\bf B565}, 269 (2000); 
P. Binetruy, {\it et al}, Phys. Lett. {\bf B477}, 285 (2000);
J. M. Cline, C. Grojean and G. Servant, 
Phys. Rev. Lett. {\bf 83}, 4245 (1999).
P. Kanti, {\it et al},  
 Phys. Lett. {\bf B468}, 31 (1999).  
R.N. Mohapatra, A. P\'erez-Lorenzana and C. A. de S. Pires, hep-ph/0003328;
C. Barcelo, M. Visser, Phys. Lett. {\bf B482}, 183 (2000);
V. Barger, T. Han, T. Li , J. D. Lykken , D. Marfatia;  hep-ph/0006275;
P. Kanti, I. I. Kogan, K. A. Olive, M. Pospelov,
Phys. Lett. {\bf B468}, 31 (1999); 
Phys. Rev. D{\bf 61}, 106004 (2000);
P. Kanti, K. A. Olive, M. Pospelov;  hep-ph/0005146.  

\bibitem{reheating1}
A. D. Dolgov and A. D. Linde;
Phys. Lett. {\bf B116}, 329 (1982);
L. F. Abbott, E. Farhi, M. B. Wise;
Phys. Lett. {\bf B117}, 29  (1982).

\bibitem{reheating2}
L. Kofman, A. Linde and  A. A. Starobinsky, 
Phys. Rev. D{\bf 56}, 3258 (1997).

 \bibitem{preheating}
L. Kofman, A. Linde and A. A. Starobinsky; 
Phys. Rev. Lett. {\bf 73}, 3195 (1994). 

\end{thebibliography}
\end{document}